
%
\magnification=\magstep1
\def\pd{\partial}
\def\log{\hbox{log}}
\def\a{\alpha}

\rightline{DTP-92/47}
\rightline{NI-92/011}
\rightline{December, 1992}

\vskip 2 true cm
\centerline{LINEARISATION OF UNIVERSAL FIELD EQUATIONS}
\vskip 2.5 true cm
\centerline{David B. Fairlie\footnote{$\sp{\dag}$}{On Research Leave from the
University of Durham}}
\vskip 0.5 true cm
\centerline{\vbox{\hbox{Department of Mathematical Sciences,}
                  \hbox{University of Durham,}
                  \hbox{\sevenrm{DURHAM DH1 3LE.}}
                   \hbox{}
                  \hbox{and}
                  \hbox{}
                  \hbox{Isaac Newton Institute}
                  \hbox{\sevenrm{CAMBRIDGE CB3 0EH.}}}}
\vskip 0.5 true cm
\centerline{\it and}
\vskip 0.5 true cm
\centerline{Jan Govaerts}
\vskip 0.5 true cm
\centerline{\vbox{\hbox{Institut de Physique Nucl\'eaire,}
                  \hbox{Universit\'e Catholique de Louvain,}
                  \hbox{2, Chemin du Cyclotron,}
                  \hbox{B - 1348 {\sevenrm{LOUVAIN-LA-NEUVE}} (Belgium).}}}
\vskip 1 true cm
\centerline{\bf Abstract}
\vskip 0.5 true cm

The Universal Field Equations, recently constructed as examples of higher
dimensional dynamical systems which admit an infinity of inequivalent
Lagrangians are shown to be linearised by a Legendre transformation. This
establishes the conjecture that these equations describe integrable systems.
While this construction is implicit in general, there exists a large class of
solutions for which an explicit form may be written.\vfill\eject
\smallskip
\centerline{\bf 1. Nonlinear Equations with an Infinity of Conservation Laws.}
\bigskip
In a  recent series of papers$\sp{[1,2,3]}$ an investigation of potentially
integrable systems in higher dimensions was initiated. The characteristic
property of the equations exhibited in those papers is that they arise
as the Euler variational equations of an infinite number of inequivalent
Lagrangians, which, since they do not involve the fields explicitly, could
be construed as providing an infinite number of conservation laws.  In the
simplest case of just one field, the equation of motion possesses the property
of covariance, i.e. any function of a solution is also a solution.

One of the most remarkable properties of this equation which led us to describe
it as
`Universal' is that it arises from an arbitrary function ${\cal F}(\phi_i),$
homogeneous of degree one in the first derivatives
$ \phi_i={\pd \phi \over\pd x_i}$ of a scalar field $\phi(x_i)$  over a
manifold of dimension $d$ by an iterative procedure of the following nature.
Denote by ${\cal E}$ the Euler differential operator
$${\cal E}=-{\pd\over\pd\phi}
 +\pd_i {\pd\over\pd\phi_i}-\pd_i\pd_j{\pd\over\pd\phi_{ij}}\dots
\eqno(1.1)$$
(In principle the expansion continues indefinitely  but it is sufficient here
 to terminate at the stage of second derivatives  $\phi_{ij}$).

Now consider the sequence of iterations;
$$\eqalign{ {\cal E F} &,\cr
            {\cal E F}{\cal E F} &,\cr
{\cal E F}{\cal E F}{\cal E F} &\quad\hbox{etc.}\cr}\eqno(1.2)$$
This sequence terminates after $d$ iterations  by vanishing identically. At the
penultimate step the resulting expression set to zero may be regarded as a
universal equation of motion; i.e. it is independent of the details of
${\cal F}$, and is in fact the equation of motion for the Lagrangian
$${\cal L}={\cal F}{\cal E F}\cdots{\cal E F}\quad (d-1\ \hbox{factors}).
\eqno(1.3)$$
This result is subject to the provision that $\cal F$ is generic, i.e that
the Hessian $M_{ij}=  {\pd\sp2{\cal F}(\phi_k)\over\pd\phi_i\pd\phi_j} $ is of
maximal rank, namely $d-1$.
In fact as it is shown in [3] it is even possible to choose different ${\cal
F}$'s in each factor in (1.3) without affecting the universality of the
resulting equation, which takes the form
$$\det\pmatrix{0&\phi_1&\phi_2&\ldots&\phi_d\cr
               \phi_1&\phi_{11}&\phi_{12}&\ldots&\phi_{1d}\cr
               \phi_2&\phi_{12}&\phi_{22}&\ldots&\phi_{2d}\cr
               \vdots&\vdots&\vdots&\ddots&\vdots\cr
               \phi_d&\phi_{1d}&\phi_{2d}&\ldots&\phi_{dd}\cr}=0.\eqno(1.4)$$
Since it arises as the Euler variation of an infinite number of Lagrangians
(1.3), it  possesses an infinite number of conservation laws, a property which
led us to speculate that this equation might be completely integrable.
A more symmetric form of this equation, suitable for further generalisation to
the multifield case can be obtained by introducing an additional variable
$x_0$, and re-defining $\phi(x_0,x_i)$ as $x_0\phi(x_i)$. Then (1.4) is
equivalent to
$$\det({\pd\sp2\phi\over\pd x_i\pd x_j}) =0\quad i,j=0,1,\dots,d.\eqno(1.5)$$
The primary objective of this paper is to demonstrate that this is indeed the
case, by exhibiting a linearisation of (1.4) using the Legendre
Transform\rlap.$\sp{[4]}$ This is a transformation which replaces a
description of a hypersurface in terms of points by a description in terms of
parameters of tangent hyperplanes, and as such is a version of Penrose's well
known twistor transform\rlap.$\sp{[5]}$
It has been used in this capacity to linearise the Plebanski equation$\sp{[6]}$
and also to construct Hyperk\"ahler manifolds\rlap.$\sp{[7]}$
\vskip 10pt
\centerline{\bf2. The Legendre Transformation }
\vskip 10pt
Suppose a scalar field $\phi(x_i)$ in $d$ dimensional space-time obeys the
dynamical equation $f(\phi(x_i),\phi_j,\phi_{jk})=0$, where subscripts denote
partial
derivatives as above. This equation can sometimes be simplified by the use of
the Legendre transformation\rlap,$\sp{[4]}$ which involves the introduction of
a set of variables $\xi_i,\ w(\xi_i)$ dual to $x_i,\ \phi(x_i)$, in the
following sense. Consider for simplicity the case $d=2$. Then $z=\phi(x,y)$
determines a surface with tangent plane at the point
$x_0,y_0,z_0=\phi(x_0,y_0)$ given by the equation
$$z-z_0 -(x-x_0)\phi_x(x_0,y_0)-(y-y_0)\phi_y(x_0,y_0)=0.\eqno(2.1)$$
Now the general equation of a plane may be specified by three parameters
$w,\xi,\eta$ as follows
$$z-\xi x-\eta y + w = 0.\eqno(2.2)$$
Comparing (2.1) and (2.2) it is evident that the conditions such that (2.2) is
a tangent plane to the surface at the point $x_0,y_0,z_0$
are
$$\xi=\phi_{x_0},\quad\eta =\phi_{y_0},\quad
w=x_0\phi_{x_0}+y_0\phi_{y_0}-z_0.
\eqno(2.3)$$
Now the surface $z=\phi(x,y)$ is also determined if $w$ is given as a function
of $\xi,\eta$ by which the two parameter family of tangent planes is
characterized. So, since $x_0,y_0,z_0$ is a generic point on the surface, we
can drop the subscript and write the conditions as
$$ \eqalign{\phi(x,y)+w(\xi,\eta)=&\ x\xi+y\eta,\cr
\xi= \phi_x ,\quad&\quad\eta=\phi_y,\cr
x=w_{\xi},\quad& \quad y=w_{\eta}.\cr}\eqno(2.4) $$
The last two relations may be obtained by partial differentiation of the first
equation with the aid  of the second two.
 This set then demonstrates a duality in the alternative descriptions of the
geometry of the situation in terms of point and plane coordinates and (2.4)
assigns to every surface element $x,y,\phi,\phi_x,\phi_y$ a surface element
$\xi,\eta,w,w_{\xi},w_{\eta}$.   This transform, which is clearly involutive
has the flavour, as was remarked earlier of a twistor transform. The
generalisation to an arbitrary number of independent variables is immediate;
$$\eqalign{\phi(x_1,x_2,\dots,x_d)+w(\xi_1,\xi_2,\dots,\xi_d)=
&x_1\xi_1+x_2\xi_2+\dots ,x_d\xi_d.\cr
\xi_i={\pd\phi\over\pd x_j},\quad x_i={\pd w\over\pd \xi_i},\quad&
\forall{i}.\cr}
\eqno(2.5)$$
To evaluate the second derivatives $\phi_{ij}$ in terms of derivatives of $w$
it is convenient to introduce two Hessian matrices;
$\Phi,\  W$ with matrix elements  $\phi_{ij}$ and $w_{\xi_i\xi_j}=w_{ij} $
respectively. Then assuming that $\Phi$ is invertible, $\Phi W=1\!\!1$
and
$$ {\pd\sp2\phi\over\pd x_i\pd x_j}= ( W\sp{-1})_{ij},\quad
 {\pd\sp2w\over\pd \xi_i\pd \xi_j}= (\Phi\sp{-1})_{ij}.\eqno(2.6)$$
Since $\det\Phi\neq 0$, equation (1.4) can be written as
$$    \sum_{i,j}\phi_i(\Phi\sp{-1})_{ij}\phi_j=0.\eqno(2.7)$$
 The effect of the Legendre transformation is immediate; in the new variables
the equation becomes simply
$$\sum_{i,j}\xi_i \xi_j{\pd\sp2w\over\pd \xi_i\pd \xi_j} =0,\eqno(2.8)$$
a linear second order equation for the function $w$! Indeed, in terms of
variables $y_j=\log(\xi_j)$
the equation is linear with constant coefficients, and is thus completely
understood. The general solution to this equation is simply
$$w(\xi_i) = v_0(\xi_i)+v_1(\xi_i),\eqno(2.9)$$
where $v_0,\ v_1$ are two arbitrary functions, homogeneous of degree zero and
one respectively in $\xi_i$.
 The solution of the original problem is given implicitly by
elimination of $\xi_i$ from the equations
$${\pd w\over\pd\xi_i}=x_i,\quad \phi(x_k)=\sum_j
x_j\xi_j-w(\xi_k)=-v_0(\xi_k).\eqno(2.10)$$
Note that the property of covariance of the solution $\phi$ is a reflection of
the arbitrariness of $v_0(\xi_k)$.
An explicit solution will not be possible in general, though in particular
cases it might be feasible. Note that equations (2.9,2.10) imply
$$\sum_j x_j{\pd \phi(x_i)\over\pd x_j}=v_1(\xi_i).\eqno(2.11)$$
Now suppose $v_1(\xi_i)=0.$
This imposes the restriction that $\phi(x_i)$ is a function, homogeneous of
degree zero in the variables $x_i$, but is otherwise arbitrary. That such an
explicit function satisfies (1.4) may be verified directly, an observation
which already has been recorded in [1]. The Legendre Transform method fails for
the choice  $v_0(\xi_i)=0.$

 As an illustrative example the case of $d=2$, the so called Bateman equation
will now be considered.
In terms of the original variables, this equation takes the form
$$ \phi_{y}\sp2\phi_{xx}-2\phi_x\phi_y\phi_{xy}+ \phi_{x}\sp2\phi_{yy}=0.
\eqno(2.12)$$
Under the Legendre transformation (2.4) this equation becomes
$$ \xi\sp2w_{\xi\xi}+2\xi\eta w_{\xi\eta}+\eta\sp2w_{\eta\eta}=0.\eqno(2.13)$$
This linear equation admits the general solution
$$w=f({\xi\over\eta})+(\xi+\eta)g({\xi\over\eta}),\eqno(2.14)$$
where $f,\ g$ are arbitrary functions. Differentiation with respect to $\xi,\
\eta$ yields
$$\eqalign{
x=w_{\xi}=&{1\over\eta}(f'+(\xi+\eta)g')+g,\cr
y=w_{\eta}=&-{\xi\over\eta\sp2}(f'+(\xi+\eta)g')+g.\cr}\eqno(2.15)$$
Thus
$$ x\xi+y\eta=(\xi+\eta)g = w+\phi,\eqno(2.16)$$
giving
$$\phi=-f({\xi\over\eta}).\eqno(2.17)$$
Thus ${\xi\over\eta}$ is an arbitrary function of $\phi$. Division of the first
relation of (2.16) by $\eta$, followed by redefinition, gives the standard
construction of a general implicit solution to the Bateman equation which runs
as follows: Constrain two arbitrary functions $f_1(\phi),f_2(\phi)$ by the
relation
$$ xf_1(\phi) +yf_2(\phi)=c\quad\hbox{(constant)},\eqno(2.18)$$
and solve for $\phi$. Then this $\phi$ solves the Bateman equation.
 For general $d$ it will not be possible to carry out the explicit
eliminination of the auxiliary variables $\xi_i$ except in very special
circumstances. The reason that this works here is that the equation (2.19) is
parabolic.

This method of solution fails when $\det\Phi=0$. A large class of evident
solutions to (1.4) fall into this category, for example those for which $\phi$
is a function of all $x_i$ except one. A less trivial example consists of those
for which  $\phi$ is given by an extension of (2.17);
$$ \sum_{i=1}\sp{i=d}x_if_i(\phi) =c \quad\hbox{(constant)},\eqno(2.18)$$
an implicit functional relation for $\phi$ in terms of arbitrary functions
$f_i(\phi)$\rlap.$\sp{[1]}$
This equation implies the following structure for the second derivatives;
$$\phi_{ij}=\a_i\phi_j +\a_j\phi_j.\eqno(2.19)$$
The precise form of the functions $\a_i,\a_j$ is not revelant, but is easily
found from (2.18). The  important point is that (2.19) implies that
$\det\Phi=0$
for $d>2$.

\vskip 10pt
\centerline{\bf3. Other Transformable Equations.}
\vskip 10pt

It is clear that many other examples of integrable nonlinear equations of
second order in field derivatives may now be constructed by  reversing the
Legendre transformation on a linear equation. Whether these equations also
enjoy similar properties to those exhibited in [1,2,3] is a matter for
speculation; however it is instructive to consider the case of the equation
$$ \phi_{t}\sp2\phi_{xx}-\phi_{x}\sp2\phi_{tt}=0,\eqno(3.1)$$
 which  results from the substitution of
$u(x,t)= -{\phi_t\over\phi_x}$ into the first order differential
equation describing nonlinear waves
$$ {\pd u\over\pd t} = u{\pd u\over\pd x}.\eqno(3.2)$$
(Substitution of $u(x,t)= + {\phi_t\over\phi_x}$ yields the Bateman equation.)
This equation can be derived from the Lagrangian
 ${\cal L}=\log {\phi_t\over\phi_x}$, and admits an infinite set of
conservation laws of the form
$$ {\pd\over\pd t}{\pd\over\pd\phi_t}F(\phi_t\phi_x)-
{\pd\over\pd x}{\pd\over\pd\phi_x}F(\phi_t\phi_x)
=0,\eqno(3.3)$$
where $F$ is an arbitrary differentiable function of the product $\phi_t\phi_x$
and $\phi(t,x)$ satisfies (3.1). Since (3.2) possesses an infinite number of
conservation laws of the form
$$  {\pd\over\pd t}(u\sp n)={\pd\over\pd x}({n\over n+1}u\sp{n+1}),\eqno(3.4)$$
where $n$ is arbitrary, there are also independent conservation laws of the
form
$$ (n+1){\pd\over\pd t}({\phi_t\over\phi_x})\sp n
+n{\pd\over\pd x}({\phi_t\over\phi_x})\sp{n+1}=0.\eqno(3.5)$$
In fact, this last equation can be written in terms of an arbitrary function
$G({\phi_t\over\phi_x})$ which admits a power series expansion as
$${\pd\over\pd t}{\pd\over\pd\phi_t}(G({\phi_t\over\phi_x})\phi_x)
-{\pd\over\pd
x}{\pd\over\pd\phi_x}(G({\phi_t\over\phi_x})\phi_x)=0.\eqno(3.6)$$
It is curious that both (3.4) and (3.6) both have the form of an Euler
variation of a Lagrangian,
 except for the introduction of a `Lorentz metric' into the Euler operator
which is here
$${\cal E}'={\pd\over\pd t}{\pd\over\pd\phi_t}
-{\pd\over\pd x}{\pd\over\pd\phi_x}\eqno(3.7)$$
and $G({\phi_t\over\phi_x})\phi_x$ is homogeneous of degree one in derivatives
of $\phi$, just like the Bateman Lagrangian.
 The equation of motion can then be written in the form
$${\cal E}'{\cal L}={\pd\over\pd t}{\pd\over\pd\phi_t}{\cal L}
-{\pd\over\pd x}{\pd\over\pd\phi_x}{\cal L}=0,\eqno(3.8)$$
with a `Lagrangian' written as
${\cal L}=F(\phi_t\phi_x)+G({\phi_t\over\phi_x})\phi_x$.
The application of the Legendre transform to (3.1) produces
the equation
$$\xi\sp2w_{\xi\xi}-\eta\sp2w_{\eta\eta}=0,\eqno(3.9)$$
with general solution
$$w=f(\xi\eta)+\eta g({\xi\over\eta}), \eqno(3.10)$$
where $f,\ g$ are arbitrary functions of one variable.
Note the appearance of a similar functional dependence to that in the
conservation laws.
The general solution of (3.1) is then obtained from the elimination of
${\xi,\ \eta}$  from the equations
$$\eqalign{
{\pd w\over\pd\xi}=\ &x=\eta f'(\xi\eta)+g'({\xi\over\eta}), \cr
{\pd w\over\pd\eta}=\ &t=\xi f'(\xi\eta)-{\xi\over\eta}g'({\xi\over\eta})+g,\cr
\phi= & 2\xi\eta f'(\xi\eta)  -f(\xi\eta),\cr }\eqno(3.11)$$
The last equation implies that the product $\xi\eta$  is an arbitrary function
of $\phi$, which might as well be taken as $\phi$ itself since equation (3.1)
possesses the same covariance property as the Universal equation, and any
function of a solution is also a solution. From the first pair of equations
(3.11) one may deduce that ${\xi\over\eta}x-t$ is an arbitrary function of the
ratio ${\xi\over\eta}$, but nothing more without making a specific  choice of
$g$.

Another example involves a slight generalisation of (1.4) to the case where
the zero in the top left corner is replaced by  $q\phi$, where $q$ is a
numerical factor. Then the equation (2.7) becomes
$$ q\phi  - \sum_{i,j}\phi_i(\Phi\sp{-1})_{ij}\phi_j=0,\eqno( 3.12)$$
which translates into
$$(\sum_j\xi_j{\pd\over\pd\xi_j})\sp2w-(q+1)\sum_j\xi_j{\pd\over\pd\xi_j}w
+qw=0,\eqno(3.13)$$
with general solution of exactly the same form as (2.9)
$$w(\xi_i) = v_q(\xi_i)+v_1(\xi_i),\eqno(3.14)$$
where $v_q,\ v_1$ are two arbitrary functions, homogeneous of degree $q$ and
one respectively in $\xi_i$, provided $q\neq1$. The solution of the original
equation proceeds in principle by elimination of $\xi_i$, using (3.13), from
$${\pd w\over\pd\xi_i}=x_i,\quad \phi(x_k)=\sum_j
x_j\xi_j-w(\xi_k)=(q-1)v_q(\xi_k). \eqno(3.15)$$ If $q=1$, then the solution
(3.14) requires modification, with attendant consequences for (3.15).
\vskip 10pt
\centerline{\bf4. Multicomponent Field Generalisation.}
\vskip 10pt
In  paper [3], a generalisation which was already conjectured in [1], of the
Universal Field Equation (1.4) to an arbitrary number  of fields, (but fewer
than the number of space-time
dimensions), was proved. Essentially the trick is to augment the number of
space co-ordinates by an additional set  $u_a$ equal to  the number $k$ of
fields $f\sp{a}(x_j)$ and write the Universal Equation (1.4) in terms of a
master field
$$\phi(u_a,x_j)=\sum_a\sp k u_af\sp{a}(x_j)\eqno(4.1)$$
Then the equation may be writtten
$$\det\pmatrix{
0&\phi_{u_1u_2}&\ldots&\phi_{u_1u_k}&\phi_{u_1x_1}&\ldots&\phi_{u_1x_d}\cr
\phi_{u_2u_1}&\phi_{u_2u_2}&\ldots&\phi_{u_ku_k}&\phi_{u_2x_1}&\ldots&
\phi_{u_2x_d}\cr
 \vdots&\vdots&\ddots& \vdots&\vdots&\ddots& \vdots\cr
\phi_{u_ku_1}&\phi_{u_ku_2}&\ldots&\phi_{u_ku_k}&\phi_{u_kx_1}&\ldots&
\phi_{u_kx_d}\cr
\phi_{x_1u_1}&\phi_{x_1u_2}&\ldots&\phi_{x_1u_k}&\phi_{x_1x_1}&\ldots&
\phi_{x_1x_d}\cr
\vdots&\vdots&\ddots& \vdots&\vdots&\ddots& \vdots\cr
\phi_{x_du_1}&\phi_{x_du_2}&\ldots&\phi_{x_du_k}&\phi_{x_dx_1}&\ldots&
\phi_{x_dx_d}\cr}=0.\eqno(4.2)$$

In fact, though the equations have been expressed in this way to
emphasise that they are effectively just a particular case of (1.5), when the
linear dependence of $\phi$ on $u_a$ is invoked, the property
$\phi_{u_au_b}=0,\ \forall a,b$ implies that the the leading $k\times k$
submatrix in (4.2) vanishes, and the first row and column may be re-expressed
after re-organising the determinant as
 $(0,\cdots,0,{\pd\phi\over\pd x_1},\cdots,{\pd\phi\over\pd x_d})$. The
coefficients of monomials of total degree $d+k$ in the variables $u_a$ in the
expansion of the left hand side of (4.2), set individually to zero form an
overdetermined set of equations for the description of the fields
$f\sp{a}(x_j)$. This set of equations is generally covariant, i.e. any set of
functions of the  solution set is also a solution set.

It is now comparatively easy to adapt the method of section (2) to the solution
of this equation. All that is necessary is to introduce a set of $k-1$
variables $\lambda_a$, conjugate to the $u_a,\ a=2,\dots ,k$ in the same way as
the $\xi_j$ are to the $x_j$. $u_1$ plays a special role, and is not
conjugated.
The Legendre Transform becomes
$$\eqalign{\phi(u_a,\ x_j)+w(\lambda_a,\ \xi_j)=&\sum_2\sp {k}u_a \lambda_a
+\sum_1\sp d x_j\xi_j,\cr
\lambda_a={\pd\phi\over\pd u_a},\quad
 u_a={\pd w\over\pd \lambda_a},\quad& \forall{a\neq 1},\cr
\xi_i={\pd\phi\over\pd x_j},\quad x_i={\pd w\over\pd \xi_i},\quad&
\forall{i}.\cr}\eqno(4.3)$$
The transform of equation (4.2) is simply (2.8) with modified solution
$$w(\lambda_a,\xi_i) = v_0(\lambda_a,\xi_i)+v_1(\lambda_a,\xi_i),\eqno(4.4)$$
where $v_0,\ v_1$ are two arbitrary functions, homogeneous of degrees zero and
one respectively
in $\xi_i$, but with thus far unrestricted dependence on $\lambda_a$. The
arbitrariness in dependence on $\lambda$ is a reflection of the general
covariance of the solution for $f\sp a.$ The information that $\phi$ is a
linear form in the variables $u_a$ must now be imposed upon the implicit
solution of the functional relationships (4.3).

The question of the introduction of this constraint complicates the issue as to
whether this is a genuinely linearisable  problem since
the conditions $\phi_{u_au_b}=0$ translate into highly nonlinear
restrictions on $W$, namely that the corresponding matrix elements of
$W\sp{-1}$ vanish.

The class of explicit solutions in section 2, may however be trivially extended
to the multifield case. All that is necessary is to observe that the choice
of $f\sp{a}(x_i)$ as a set of arbitrary functions, homogeneous of degree zero
in their arguments automatically satisfies (4.2)!

\vskip 10pt
\centerline{\bf5. Conclusion.}
\vskip 10pt
This analysis has demonstrated that the Universal Field Equations proposed
in [1,2] which are  covariant in the field, or reparametrisation invariant in
the base space are linearisable by a Legendre Transform, and thus may be added
to the dossier of examples of integrable systems linearisable by a transform
method.  It thus justifies the hopes for integrability presented in those
papers, based upon the existence of an infinite number of conservation laws.
(There was a flurry of activity in the mid 60's
when such conservation laws were written down for linear
systems)\rlap.$\sp{[8,9,10,11]}$

The more general class of overdetermined equations, conjectured in [1,2] and
demonstrated to be simply a particular case of the single field in $d+k-1$
dimensions, and thus potentially integrable, still require a further technical
trick before the assertion of integrability can be justified, on account of the
difficulty in implementing the requirement of a linear decomposition of $\phi$
into its component fields.
  The most interesting case  is that for $d=4,\ k=2$. Then the equations
describe a new reparametrisation invariant string in 4 dimensions,
whose world sheet is specified by the intersection of the hypersurfaces
$f(x,y,z,t)=0$ and $g(x,y,z,t)=0$.
More explicitly this set of equations is given by requiring
$$\det\pmatrix{
0&0&f_x&f_y&f_z&f_t\cr
0&0&g_x&g_y&g_z&g_t\cr
f_x&g_x&u_1f_{xx}+u_2g_{xx}&u_1f_{xy}+u_2g_{xy}&u_1f_{xz}+u_2g_{xz}
&u_1f_{xt}+u_2g_{xt}\cr
f_y&g_y&u_1f_{xy}+u_2g_{xy}&u_1f_{yy}+u_2g_{yy}&u_1f_{yz}+u_2g_{yz}
&u_1f_{yt}+u_2g_{yt}\cr
f_z&g_z&u_1f_{xz}+u_2g_{xz}&u_1f_{yz}+u_2g_{yz}&u_1f_{zz}+u_2g_{zz}
&u_1f_{zt}+u_2g_{zt}\cr
f_t&g_t&u_1f_{xt}+u_2g_{xt}&u_1f_{yt}+u_2g_{yt}&u_1f_{zt}+u_2g_{zt}
&u_1f_{tt}+u_2g_{tt}\cr}=0\eqno(5.1)$$
for all choices of $ u_1,\  u_2.$ It will be interesting to examine solutions
either based upon the techniques of this paper, or otherwise. Note the
particularly simple class of solutions $f(x,y,z,t),\ g(x,y,z,t)$ as arbitrary
homogeneous functions of degree zero in $ x,y,z,t$. The corresponding
hypersurfaces mentioned above are now generalised cones.

For a large class of known solutions  (2.18), the Legendre Transform method
technically fails, because $\Phi$ is singular for these solutions. Furthermore,
although these equations are linearisable by the Legendre Transform, this does
not mean that they are tractable, as the solution so obtained is only implicit.
This is by no means unusual in dealing with nonlinear systems; for example the
well known ADHM construction$\sp{[12]}$ of  $SU(2)$ self dual Yang Mills
instantons is
not explicitly solvable in the general case, nor is the corresponding
Multi-monopole construction\rlap.$\sp{[13]}$

The intriguing feature of the Universal Field Equations is that they are
derivable by a process of iteration of the Euler variation, for which there is
not so far a geometrical motivation. By applying the converse transformation to
other linear systems, such as those exemplified in section 3, more
understanding of the nature of the higher dimensional integrable systems of
this type may be gained.

\vfill\eject
\vskip 10pt
\centerline{\bf REFERENCES}
\frenchspacing
\item{[1]}D.B. Fairlie, J. Govaerts and A. Morozov, {\it  Nuclear Physics}
\ {\bf  B373}\ (1992)\ 214-232.
\item{[2]}D.B. Fairlie, J. Govaerts, {\it Physics Letters}\ {\bf 281B}\ (1992)
49-53.
\item{[3]}D.B. Fairlie, J. Govaerts, {\it J. Math Phys}{\bf 33}\ (1992)\
3543-3566.
\item{[4]}R. Courant and D. Hilbert, {\it Methods of Mathematical Physics Vol
II}
Interscience, (1962)\ 32-39.
\item{[5]}R.Penrose `Twistor Theory; its aims and achievements' {\it Quantum
Gravity} Eds C.J. Isham, R. Penrose and D. Sciama O.U.P. (1975) 268-407.
\item{[6]}C.M. Hull, ${\cal W}$- Geometry, Queen Mary and Westfield College
preprint (QMW-92-6) 1992.
 \item{[7]}N. Hitchin, A. Karlhede, U. Lindstrom and M. Ro\v cek, {\it Comm
Math Phys.}\  {\bf 108}\ (1987)\  535-589.
\item{[8]}D.J. Candlin, {\it Il Nuovo Cimento}\ {\bf 37}\ (1965)\ 390-395.
\item{[9]}D.B. Fairlie, {\it Il Nuovo Cimento}\ {\bf 37}\ (1965)\ 897-904.
\item{[10]}D.M. Fradkin, {\it J. Math. Phys.}\ {\bf 5}\ (1965)\ 879-890.
\item{[11]}T.W.B. Kibble, {\it J. Math. Phys.}\ {\bf 5}\ (1965)\ 879-890.
\item{[12]} M.F. Atiyah, N.J. Hitchin, V.G. Drinfeld and Yu.I. Manin, {\it
Physics Letters}\ {\bf A65}\ (1978)\  185-187.
\item{[13]}E.F. Corrigan and P. Goddard, {\it Comm Math. Phys.}\ {\bf 80}\
(1981)\
 575-587.

\vfill\eject
\end